\documentclass[aps,prl,superscriptaddress,twocolumn,showpacs,titlepage=false]{revtex4-1}
\usepackage{graphicx}
\usepackage{amsmath}
\usepackage{amssymb}
\usepackage{subfigure}
\usepackage{natbib}
\usepackage{color}
\usepackage{blkarray}
\usepackage{dsfont}
\usepackage{MnSymbol}
\usepackage[breaklinks=true,colorlinks=true,linkcolor=blue,urlcolor=blue,citecolor=blue]{hyperref}
\usepackage[percent]{overpic}
\usepackage{mathrsfs}
\usepackage{verbatim}
\usepackage{bbold}
\usepackage{rotating}
\def\bra#1{{\left\langle #1 \right|}}
\def\ket#1{{\left| #1 \right\rangle}}

\newcommand{\bs}[1]{\boldsymbol{#1}}
\def\Id{\mathds{1}}
\newcommand{\Tr}{\mathrm{Tr}}

\begin{document}
\title{What randomized benchmarking actually measures}
\author{Timothy Proctor}
\affiliation{Sandia National Laboratories, Livermore, CA 94550, USA}
\author{Kenneth Rudinger}
\affiliation{Center for Computing Research, Sandia National Laboratories, Albuquerque, NM 87185, USA}
\author{Kevin Young}
\author{Mohan Sarovar}
\affiliation{Sandia National Laboratories, Livermore, CA 94550, USA}
\author{Robin Blume-Kohout}
\affiliation{Center for Computing Research, Sandia National Laboratories, Albuquerque, NM 87185, USA}
\date{\today}

\begin{abstract}
Randomized benchmarking (RB) is widely used to measure an error rate of a set of quantum gates, by performing random circuits that would do nothing if the gates were perfect. In the limit of no finite-sampling error, the exponential decay rate of the observable survival probabilities, versus circuit length, yields a single error metric $r$.  For Clifford gates with arbitrary small errors described by process matrices, $r$ was believed to reliably correspond to the mean, over all Cliffords, of the \emph{average gate infidelity} (AGI) between the imperfect gates and their ideal counterparts.  We show that this quantity is not a well-defined property of a physical gateset.  It depends on the \emph{representations} used for the imperfect and ideal gates, and the variant typically computed in the literature can differ from $r$ by orders of magnitude.  We present new theories of the RB decay that are accurate for all small errors describable by process matrices, and show that the RB decay curve is a simple exponential for all such errors. These theories allow explicit computation of the error rate that RB measures ($r$), but as far as we can tell it does not correspond to the infidelity of a physically allowed (completely positive) representation of the imperfect gates.
\end{abstract}
\maketitle

Randomized benchmarking (RB) \cite{emerson2005scalable,emerson2007symmetrized,dankert2009exact,knill2008randomized,magesan2011scalable,magesan2012characterizing,wallman2014randomized,epstein2014investigating,magesan2012efficient,wallman2015estimating,kimmel2014robust,gambetta2012characterization,alexander2016randomized,helsen2017multiqubit,fogarty2015nonexponential,carignan2015characterizing,cross2016scalable,granade2015accelerated,sheldon2016characterizing,ball2016effect,kelly2014optimal,wallman2015robust} is a simple and efficient protocol for measuring an average error rate of a quantum information processor (QIP), and is among the most commonly used experimental methods for characterizing QIPs \cite{barends2014superconducting,chen2016measuring,barends2014rolling,xia2015randomized,muhonen2015quantifying,corcoles2015demonstration,laucht2015electrically,chow2014implementing,veldhorst2014addressable,johnson2015demonstration,corcoles2013process}.  In its purest form, RB consists of: (1) performing many randomly chosen sequences of Clifford gates that ought to return the QIP to its initial state; (2) measuring at the end of each sequence to see whether the QIP ``survived'' (i.e., returned to its initial state); and (3) plotting the observed survival probabilities vs. sequence length and fitting this to an exponential decay curve. The decay rate of the survival probability is -- up to a dimensionality constant, and neglecting any finite-sampling error -- the ``RB number'' ($r$).  RB experiments estimate $r$, which is used as a metric for judging the processor's performance.

\begin{figure}
\begin{overpic}[width=8.0cm]{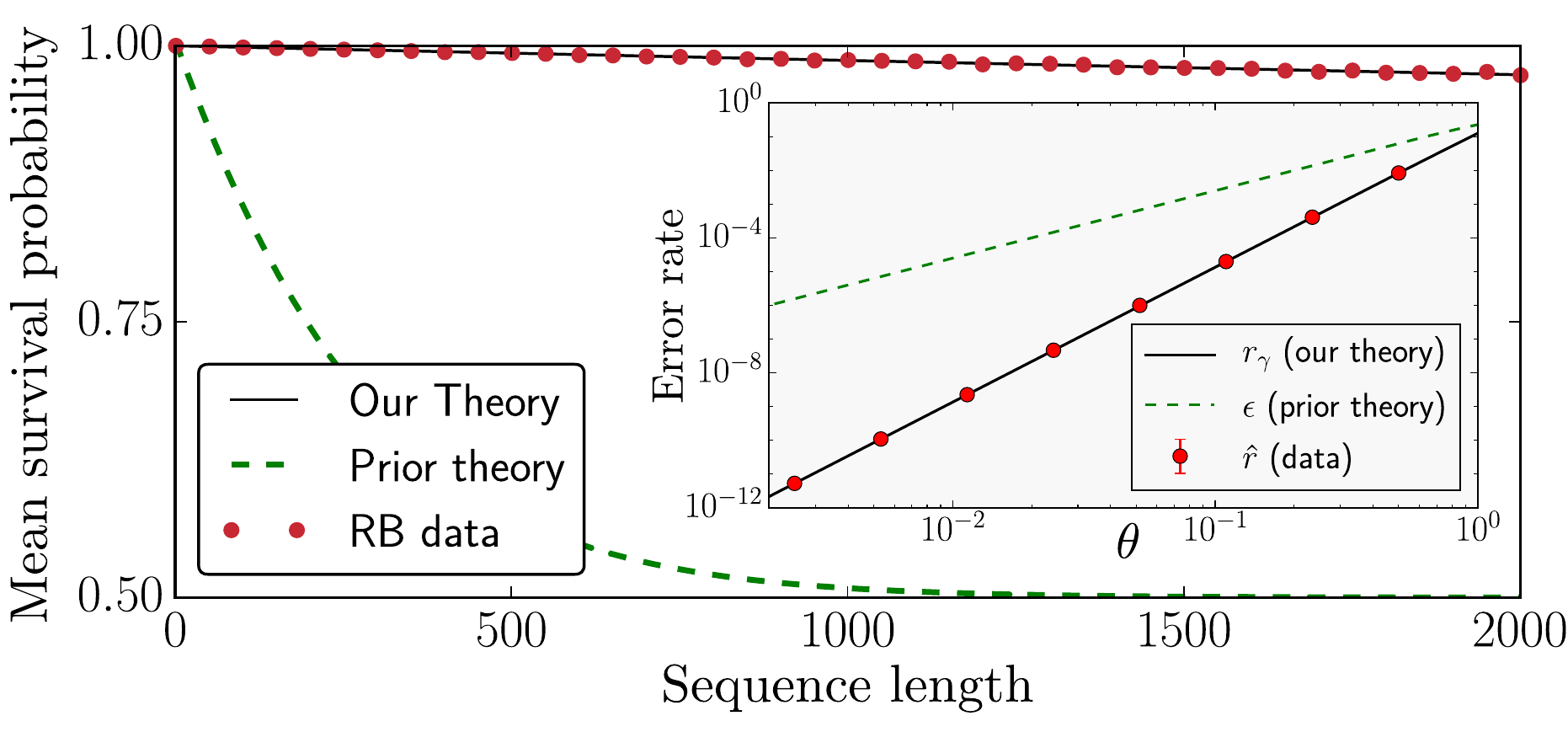}
\end{overpic}
\caption{An example of the discrepancy between $r$ and the literature definition for $\epsilon$. Here the errors are coherent rotations proportional to $\theta$ (see main text). For $\theta \ll 1$ the errors are small, and prior RB theory \cite{magesan2012characterizing,magesan2011scalable,wallman2014randomized,epstein2014investigating} predicts that $r \approx \epsilon$. Main plot: the curve predicted by prior RB theory \cite{magesan2012characterizing,magesan2011scalable} (for $\theta = 0.1$) is inconsistent with simulated RB data, which is accurately predicted by the theory introduced herein. Inset: $\epsilon$, estimates of $r$ from simulated data, and the $r$ predicted by our theory ($r_{\gamma}$), as $\theta$ is varied.}
\label{Fig:scaling}
\end{figure}

The $r$ that RB measures has a clear operational definition, but it is not clear how it relates to common metrics -- i.e., what it is that RB measures. In QIP theory, the ideal ``target'' operations and the imperfect as-implemented operations are usually represented by \emph{process matrices}, a.k.a. CPTP (completely positive, trace-preserving) maps. The generally accepted theory behind RB \cite{magesan2011scalable,magesan2012characterizing,wallman2014randomized,epstein2014investigating} suggests that $r$ is approximately equal to the average, over all $n$-qubit Cliffords, of the \emph{average gate infidelity} [AGI, Eq.~\eqref{eq:AGF}] between the imperfect Cliffords and their ideal counterparts.  We call this quantity the \emph{average gateset infidelity} [AGsI, Eq.~\eqref{eq:AGsI}] and denote it by $\epsilon$. It has been widely believed that $r \approx  \epsilon$ whenever the errors in the gates are small, and describable by process matrices \cite{magesan2011scalable,magesan2012characterizing,wallman2014randomized,epstein2014investigating,magesan2012efficient,
wallman2015estimating,kimmel2014robust,gambetta2012characterization,alexander2016randomized,helsen2017multiqubit,fogarty2015nonexponential,carignan2015characterizing,cross2016scalable}. In this Letter, we show that $r$ and $\epsilon$ can differ by orders of magnitude  (Fig.~\ref{Fig:scaling}). This happens because $\epsilon$ is not a well-defined property of a physical QIP.  Instead, $\epsilon$ is a property of the \emph{representation} used to describe the gates, and depends strongly on which of several equivalent and indistinguishable representations is used.  We provide a new theory for the RB decay that \emph{is} representation-independent, proves that the RB decay is always exponential when the noise is described by process matrices, and gives an efficient representation-independent approximate formula for $r$ with small error bars.

\vspace{0.1cm}
\noindent\textbf{Experimental RB:}  The basic RB protocol (extensions exist \cite{magesan2012efficient,kimmel2014robust,gambetta2012characterization,wallman2015estimating,alexander2016randomized}) was summarized above. Complete details can be found in Refs.~\cite{magesan2011scalable,magesan2012characterizing,wallman2014randomized,epstein2014investigating}, and in the appendix. As in most experiments \cite{barends2014superconducting,chen2016measuring,barends2014rolling,xia2015randomized,muhonen2015quantifying,corcoles2015demonstration,laucht2015electrically,chow2014implementing,veldhorst2014addressable}, we consider benchmarking an implementation of the $n$-qubit Clifford gates with $n \geq 1$.  The standard way to estimate $r$ from RB data is to fit the average of the sampled survival probabilities ($P_m$), for many sequence lengths $m$, to the model $P_m = A + (B + Cm)p^m$, where $A$, $B$, $C$, and $p$ are fit parameters  \cite{magesan2011scalable,magesan2012characterizing,wallman2014randomized,epstein2014investigating}.  The estimate of $p$, denoted $\hat{p}$, gives an estimate of $r$ as $\hat{r} = (d - 1)(1 - \hat{p})/d$, where $d = 2^n$.  It is common to fix $C = 0$, but Magesan \emph{et al.}~\cite{magesan2012characterizing,magesan2011scalable} suggest that fitting $C$ may be necessary when the error varies from gate to gate. 

\vspace{0.1cm}
\noindent\textbf{Theory of RB:} The average survival probabilities $P_m$ are unambiguously real and experimentally accessible.  And $r$ is equally well-defined, as long as the $P_m$ decay exponentially with $m$.  The motivation for further analysis -- for a \emph{theory} of RB -- is primarily to answer two questions.  First, under what circumstances does $P_m$ decay exponentially?  Second, when it does, what is $r$?  That is, to what property of the imperfect gates does $r$ correspond?  Building such a theory requires specifying a model for the operations used in RB.

These operations comprise: (1) a set of gates; (2) a set of state preparations; and (3) a set of measurements, which together form a \emph{physical gateset}. A model associates them with mathematical objects that can be used to compute $P_m$.  If each operation is independent of all external contexts -- e.g., time, external fields, ancillary qubits -- then each gate can be represented by a \emph{process matrix} $G_i$, each state preparation by a density operator $\rho_j$, and each measurement by a positive operator-valued measure (POVM) $\mathcal{M}_{k} = \{ E_{k,l} \}$.  Probabilities of events are given by Born's Rule:  $\mathrm{Pr}(E_{k,l}\, | \,\rho_j, G_i) = \Tr\left[E_{k,l} G_i \rho_j\right]$. In this commonly used model for analyzing RB, an as-built processor with an imperfect physical gateset can be \emph{represented} by some $\tilde{\mathcal{G}}  = \{ \tilde{G}_i, \tilde{\rho}_j , \tilde{E}_{k,l} \}$, and an idealized perfect device by some $\mathcal{G}  = \{ G_i, \rho_j , E_{k,l} \}$. Since $r$ is independent of the state preparation and measurement \cite{magesan2012characterizing,magesan2011scalable}, we will usually only need representations of the imperfect and ideal Cliffords, denoted $\tilde{\mathcal{C}} = \{\tilde{C}_i\}$ and $\mathcal{C} =\{C_i\}$, respectively.

RB theory is clear when the gateset has \emph{gate-independent errors}; which means that there is a process matrix $\Lambda$ such that each imperfect Clifford can be represented as $\tilde{C}_i = \Lambda C_i$.  In this situation, $r$ is exactly equal to the \emph{average gate infidelity} (AGI) between $\Lambda$ and the identity process matrix $\Id$ \cite{magesan2011scalable}.  The AGI between process matrices $\tilde{G}$ and $G$ is simply $1 - \bar{F}$, where
\begin{equation}
\bar{F}(\tilde{G},G) := \int d\psi \,\, \text{Tr} \left(\tilde{G} [|\psi \rangle \langle \psi |] G[|\psi \rangle \langle \psi | ] \right).
\label{eq:AGF}
\end{equation}
But a general theory of RB needs to address the more likely case of gate-\emph{dependent} errors, where $\tilde{C}_i = \Lambda_i C_i$.  A starting point is the observation that, for gate-independent errors, every imperfect Clifford has the same AGI with its ideal counterpart:  $\bar{F}(\tilde{C}_i, C_i) = \bar{F}(\Lambda, \Id)$.  So, a plausible generalization of AGI to gate-dependent errors is its \emph{average} over all Cliffords:
\begin{equation}
\epsilon( \tilde{\mathcal{C}},\mathcal{C} ) := \text{avg}_{i}\left[ 1 - \bar{F}(\tilde{C}_i,C_i) \right],
\label{eq:AGsI}
\end{equation}
a quantity we call the \emph{average gateset infidelity} (AGsI). 

An extensive literature suggests or argues that $r \approx \epsilon$ \cite{magesan2011scalable,magesan2012characterizing,wallman2014randomized,epstein2014investigating,magesan2012efficient,wallman2015estimating,kimmel2014robust,gambetta2012characterization,alexander2016randomized,helsen2017multiqubit,fogarty2015nonexponential,carignan2015characterizing,cross2016scalable} for ``weakly gate-dependent'' errors \cite{magesan2012characterizing,magesan2011scalable} -- i.e.,  when all the error maps $\Lambda_i = \tilde{C}_iC_i^{-1}$ are close to their average. More precisely, when $\delta := \|\Lambda_i -  \bar{\Lambda}\|_{1\to1}^{H} \ll 1$ for all $i$ \cite{magesan2012characterizing,magesan2011scalable}, where $\bar{\Lambda} := \text{avg}_i[\Lambda_i]$ is the \emph{average error map}, and $\| \cdot \|_{1\to1}^{H} $ is the Hermitian 1-to-1 norm \cite{magesan2012characterizing}.  Since this is true whenever the $\Lambda_i$ are all close to $\Id$, it holds for all small errors.  However, $r$ and $\epsilon$ can actually differ by orders of magnitude, for simple and realistic noise models.  Consider a simple 1-qubit example involving Cliffords compiled into two ``primitive'' gates.  

\vspace{0.1cm}
\noindent\textbf{Example 1:} The ideal primitive gates are represented by $G_x =  R(\sigma_x, \pi/2)$ and $G_y = R(\sigma_y, \pi/2)$, where $R(H, \theta)[\rho] := \exp(- i \theta H/2)\rho\exp(i \theta H/2)$. Any 1-qubit Clifford can be compiled into $G_x$ and $G_y$.  The \emph{imperfect} primitives are represented by $\tilde{G}_{x} = R(\sigma_z, \theta)G_x$ and $\tilde{G}_y = R(\sigma_z, \theta)G_y$ with $\theta \ll 1$, which corresponds to a small systematic detuning or timing error. 

We simulated RB with Cliffords compiled into these imperfect gates and observed $r \ll \epsilon$.  For $\theta = 0.1$, the theory predicts $\epsilon \approx  10^{-3}$, but we observed $\hat{r} \approx 10^{-5}$ (Fig.~\ref{Fig:scaling}).  Varying $\theta$ (Fig.~\ref{Fig:scaling}, inset) shows that $r \propto \theta^4$, while $\epsilon \propto \theta^2$.  As the errors become small, the ratio $\epsilon/r$ diverges.

\begin{figure}
\begin{overpic}[width=8.5cm]{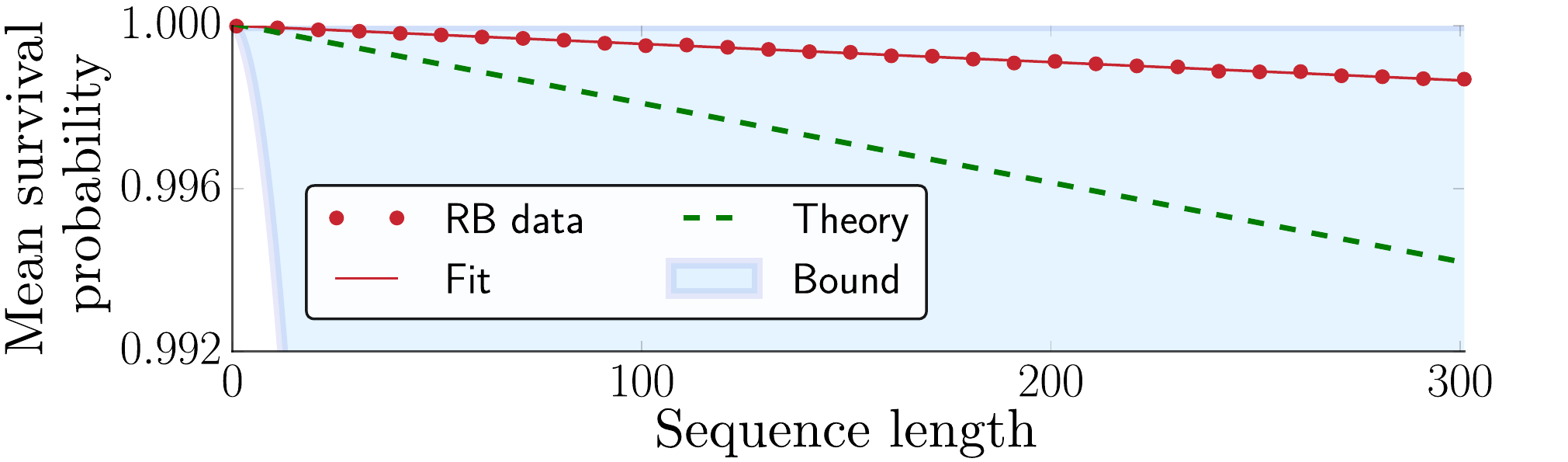}
\end{overpic}
\caption{A comparison between simulated RB data and the decay curve predicted by prior RB theory \cite{magesan2012characterizing,magesan2011scalable} for a gateset with small unitary errors. The blue shaded region depicts the range within which the RB decay is guaranteed to fall by the theorems in Ref.~\cite{magesan2012characterizing,magesan2011scalable}, in the limit of many samples.}
\label{fig:error-bounds}
\end{figure}

This example lies within the domain of standard RB theory -- the errors are small and only weakly gate-dependent (as defined in Refs.~\cite{magesan2012characterizing,magesan2011scalable}) -- and it does not contradict the technical results of Refs.~\cite{magesan2011scalable,magesan2012characterizing}, that link $r$ to $\epsilon$. Refs.~\cite{magesan2011scalable,magesan2012characterizing} include error bounds that bound the difference between actual and predicted RB decay curves. These bounds, which we plot for Example 1 in Fig.~\ref{fig:error-bounds}, are sufficiently loose that they do not significantly constrain $\epsilon/r$.  A complete description of our simulation methodology is provided in the appendix.

\vspace{0.1cm}
\noindent
\textbf{Understanding the discrepancy:}  The discrepancy between $r$ and $\epsilon$ has a simple but subtle explanation:  RB, like all experiments, probes properties of a \emph{physical} QIP, not of a \emph{model} for it.  Although a physical QIP's gates may be accurately represented by a fixed set of process matrices, that representation is not unique.  The RB error rate $r$ is a property of the physical gates, and therefore representation-independent.  But $\epsilon$, as conventionally defined, is not.

Two representations of a physical gateset are equivalent if they cannot be distinguished by any experiment.  More precisely, representations $\mathcal{G}$ and $\mathcal{G}'$ are equivalent iff they predict the same probabilities for \emph{every} quantum circuit.  Equivalent representations are easy to construct.  If $\mathcal{G} = \{G_i, \rho_j, E_{k,l}\}$ accurately represents a QIP, then so does 
\begin{equation} 
\mathcal{G}(M) = \{MG_iM^{-1}, M (\rho_j), M^{-1}(E_{k,l})\},
\end{equation}
 where $M$ is any invertible linear map, which we call a \emph{gauge transformation} \cite{blume2013robust,merkel2013self,greenbaum2015introduction,blume2016certifying}. If $f$ is an observable property of the QIP that can be computed from a model $\mathcal{G}$, then $f(\mathcal{G})$ must be the same for all equivalent representations.  So observable properties like $r$ must correspond to \emph{gauge-invariant} functions:  $f(\mathcal{G}) = f(\mathcal{G}(M))$ for all $M$.  

The AGsI defined in Eq.~(\ref{eq:AGsI}) is not gauge-invariant. It depends on the representations for the physical and perfect gatesets. If $\tilde{\mathcal{C}}$ and $\mathcal{C}$ are representations for the imperfect and ideal Cliffords respectively, then $\tilde{\mathcal{C}}(M)$ and $\mathcal{C}(N)$ are equivalent representations, for arbitrary invertible $M$ and $N$. The AGsI has a continuum of values as $M$ and $N$ are varied, and this is still true if we (arbitrarily) fix either representation.

Transforming the perfect and imperfect Cliffords in the same way (i.e., $M = N$ above) leaves the AGsI unchanged. So, we can define a gauge-invariant AGsI by comparing $\tilde{\mathcal{C}}$ \emph{not} to the usual fixed representation of the Cliffords $\mathcal{C}$, but to a $\mathcal{C}$-dependent representation of them, $\mathcal{C}_{\tilde{\mathcal{C}}}$, that satisfies $\mathcal{C}_{\tilde{\mathcal{C}}(M)} = \mathcal{C}_{\tilde{\mathcal{C}}}(M)$. For example, we could define the AGsI with respect to the representation of the perfect Cliffords that is ``closest'' to the process matrices representing the imperfect Cliffords.  If we do so, the assertion that $r \approx \epsilon$ is not \emph{wrong}, but \emph{ambiguous}; it requires a unique definition for the ``closest'' representation of the Cliffords. We return to this at the end of the Letter.

As far as we can tell, $\epsilon$ has not been defined or calculated in a representation-independent way in the literature. It is generally defined by: (1) taking $\mathcal{C}$ as the automorphism group of the Pauli matrices; (2) taking the imperfect gateset to be $\tilde{\mathcal{C}} = \{\Lambda_i C_i\}$ where the $\Lambda_i$ describe the ``relevant error process'';  and (3) calculating the AGsI (Eq.~\ref{eq:AGsI}) between $\tilde{\mathcal{C}}$ and the already-defined matrices $\mathcal{C}$.  This procedure, which we followed in our example above, is explicit in the RB simulations of Refs.~\cite{epstein2014investigating,carignan2015characterizing} and is the most natural reading of the foundational RB papers by Magesan \emph{et al.}~\cite{magesan2011scalable,magesan2012characterizing}.

\vspace{0.1cm}
\noindent\textbf{Example 2:}  A perfect Clifford gateset $\tilde{\mathcal{C}} = \mathcal{C}$ has an AGsI to $\mathcal{C}$ of $\epsilon(\tilde{\mathcal{C}},\mathcal{C}) = 0$.  But if $U$ is a unitary and $\mathcal{U}[\rho] := U\rho U^\dagger$, then $\tilde{\mathcal{C}}(\mathcal{U})$ is an equivalent representation of the gateset with generally non-zero AGsI.

Example 1 is actually very similar to Example 2.  The imperfect primitive gates in Example 1, $\tilde{G}_x$ and $\tilde{G}_y$, are \emph{almost} gauge-equivalent to their perfect counterparts. Some algebra shows that $\tilde{G}_{x/y}(\rho) = U_{x/y}\rho U_{x/y}^{\dagger}$ where $U_{x/y} = \exp( -i \phi (\hat{v}_{x/y} \cdot \vec{\sigma})/2)$, $\phi = \pi/2 + O(\theta^2)$, and $\hat{v}_x \cdot \hat{v}_y = 0 + O(\theta^2)$.  So at $O(\theta)$ the $\tilde{G}_{x}$ and $\tilde{G}_{y}$ gates induce rotations by $\pi/2$ around orthogonal axes.  Hence, there exists some $\mathcal{U}$ with $\mathcal{U}[\rho]=U\rho U^{\dagger}$ for unitary $U$ such that $\mathcal{U}\tilde{G}_{x/y}\mathcal{U}^{-1} = R(\hat{w}_{x/y} \cdot \vec{\sigma},\varphi_{x/y})G_{x/y}$ where $\varphi_{x/y} = O(\theta^2)$ and $\hat{w}_{x/y}$ are some unit vectors.  In this representation, the Clifford error maps $\Lambda_i=\tilde{C}_iC_i^{-1}$ are unitary rotations by $O(\theta^2)$, which suggests an RB number of $r = O(\theta^4)$, as observed. Although the $O(\theta)$ detuning error is real and physical, its effect on \emph{these} gates is, at $O(\theta)$, equivalent to a gauge transformation. So, in \emph{all} circuits consisting of only these gates, it behaves like a coherent error with a rotation angle of $O(\theta^2)$.

\vspace{0.1cm}
\noindent\textbf{New theories for the RB decay:}  We would like to know what property of a physical gateset RB \emph{is} measuring, and to have an accurate, efficient formula for $r(\{\tilde{C}_i\})$.  To this end, we now present new theories for the RB decay that are representation-independent and highly accurate.  

The average survival probability over all RB sequences of length $m$ is
\begin{equation}
P_m = \frac{1}{|\mathcal{C}|^m}\sum_{\bs{s}}  \text{Tr} (E \tilde{C}_{\bs{s}^{-1}} \tilde{C}_{s_m} \dots \tilde{C}_{s_{1}}  (\rho )) ,
\label{Pmdef}
\end{equation}
 where $E$ and $\rho$ are the (imperfect) measurement and state preparation, $C_{\bs{s}^{-1}}$ is the Clifford that inverts the first $m$ Cliffords, $\bs{s} \in [1..|\mathcal{C}|]^m$, and $|\mathcal{C}|$ is the order of the Clifford group. The map $\mathcal{S}_m = \text{avg}_{\bs{s}}[ \tilde{C}_{\bs{s}^{-1}} \tilde{C}_{s_m} \dots \tilde{C}_{s_{1}}]  $ can be written as $\mathcal{S}_m = | \mathcal{C}| \vec{v}^T \mathscr{R}^{m+1} \vec{v}$, where $\vec{v} = (\mathbb{1},\mathbb{0}, \dots, \mathbb{0} )^T$, $\mathbb{1}$ and $\mathbb{0}$ are the $n$-qubit identity and ``zero'' superoperators ($\mathbb{0}(\rho) = 0$) respectively, and
\begin{equation}
\mathscr{R} = \frac{1}{|\mathcal{C}|}\begin{pmatrix}
  \tilde{C}_{1 \to 1} &   \tilde{C}_{2 \to 1}  & \cdots & \tilde{C}_{|\mathcal{C}| \to 1} \\
   \tilde{C}_{1 \to 2}  &   \tilde{C}_{2 \to 2}  & \cdots & \tilde{C}_{|\mathcal{C}| \to 2} \\
  \vdots  & \vdots  & \ddots & \vdots  \\
  \tilde{C}_{1 \to |\mathcal{C}|} & \tilde{C}_{2 \to |\mathcal{C}|} & \cdots & \tilde{C}_{|\mathcal{C}|\to |\mathcal{C}|}
 \end{pmatrix},
\end{equation}
where $C_{j\to k} = C_{j}^{-1}C_k$ and $\tilde{C}_{j \to k}$ is the corresponding imperfect Clifford. It follows that $P_m = |\mathcal{C}| \, \text{Tr}(E(\vec{v}^T \mathscr{R}^{m+1} \vec{v})(\rho))$, and so
\begin{equation}
P_m = \sum_i \alpha_i \lambda_i^{m+1},
\end{equation}
 where $\{\lambda_i\}$ are the $4^n|\mathcal{C}|$ eigenvalues of $\mathscr{R}$, $n$ is the number of qubits, and $\{\alpha_i\}$ are constants depending on $\rho$, $E$, and the eigenvectors of $\mathscr{R}$.

This exact expression for the RB decay curve can be calculated efficiently in $m$ (unlike exhaustive averaging over $|\mathcal{C}|^{m-1}$ sequences). However, it is intractable for $n > 1$ qubits, and does not explain why decays with a functional form of $A + Bp^m$ are normally observed in practice. We therefore make a small approximation.

Because $\tilde{C}_{\bs{s}^{-1}} = (\bar{\Lambda} + \Delta_{\bs{s}^{-1}})  C_{s_1}^{-1} \dots C_{s_m}^{-1}$, where $\Delta_{i} = \Lambda_i - \bar{\Lambda}$, we can rewrite Eq.~(\ref{Pmdef}) as
\begin{equation}
P_m = \frac{1}{|\mathcal{C}|^m} \sum_{\bs{s}} \text{Tr} (E \bar{\Lambda}   C_{s_1}^{-1} \dots C_{s_m}^{-1} \tilde{C}_{s_m} \dots \tilde{C}_{s_{1}} (\rho)) +\tilde{\delta}_m.
\end{equation}
Therefore $P_m = \text{Tr}(E \bar{\Lambda} [\mathscr{L}^m(\Id)](\rho )) + \tilde{\delta}_m$, where
\begin{equation}
\mathscr{L}(\mathcal{E}) = \text{avg}_i[C_{i}^{-1} \mathcal{E} \tilde{C}_{i} ],
\end{equation}
 is a linear map on superoperators (a ``superduperoperator'' \cite{crooks2008quantum}; note that a similar matrix was constructed in Ref.~\cite{chasseur2015complete} to analyze leakage in RB). Hence, $P_m = \sum_i \omega_i\gamma_i^m + \tilde{\delta}_m$, where $\{\gamma_i \}$ are the $16^n$ eigenvalues of $\mathscr{L}$, and $\{\omega_i\}$ depend on $\rho$, $E$, $\bar{\Lambda}$, and the eigenvectors of $\mathscr{L}$.  The $\{\gamma_i\}$ are representation-independent. In the appendix we prove that $\tilde{\delta}_m$ satisfies $|\tilde{\delta}_m| \leq \delta_{\diamond} \equiv \frac{1}{2}\text{avg}_i \|\Lambda_i -  \bar{\Lambda}\|_{\diamond}$. Because the $\Lambda_i$ are representation-dependent, the size of $\delta_{\diamond}$ depends on the representations $\tilde{\mathcal{C}}$ and $\mathcal{C}$. But since this bound holds for \emph{any} CPTP representations, $|\tilde{\delta}_m| \leq \delta_{\diamond}^{\min}$ with $\delta_{\diamond}^{\min}$ the minimum of $\delta_{\diamond}$ over all CPTP representations of the gatesets.

If there exists a representation in which the errors are gate-independent ($\tilde{C}_i = \Lambda C_i$ for all $i$ and some $\Lambda$), then $\delta_{\diamond}^{\min} = 0$ and the RB decay is exactly described by $\mathscr{L}$. Because the Cliffords are a unitary 2-design \cite{dankert2009exact}, $\mathscr{L}$ has only three distinct eigenvalues in this case: 1, $\gamma$, and 0 (0 has a degeneracy of $16^n - 2$). The RB decay is then exactly described by $P_m = \omega_0 + \omega_1\gamma^m$. This recovers the exact RB theory for gate-independent error maps \cite{magesan2012characterizing,magesan2011scalable}.

Small errors are a small perturbation away from the case of no error ($\gamma = 1$), and cause similarly small perturbations of the eigenvalues.  Hence, for any small errors, $\gamma_0 = 1$ (as 1 is always an eigenvalue of $\mathscr{L}$), $\gamma_1$ satisfies $1 - \gamma_1 \ll 1$, and $|\gamma_{i}| \ll 1$ for all $i > 1$. As such,
\begin{equation}
P_m = \omega_0 + \omega_1 \gamma^m + \delta_m,
\end{equation}
where $\gamma = \gamma_1$, $|\delta_m| \leq \delta_{\diamond}^{\min} + \kappa_m$, and $\kappa_m = |\omega_{2}\gamma_2^m + \omega_{3}\gamma_3^m + \dots|$ is an exponentially decreasing function of $m$. Hence, for $m \gg 1$ the RB decay curve is well approximation by the functional form $P_m = A + Bp^m$. Therefore, the $p$ obtained from fitting RB data to $P_m = A + Bp^m$ is an estimate of $\gamma$, the second largest eigenvalue of $\mathscr{L}$. Similarly, as $r$ is given by $r=(d - 1)(1 - p)/d$, $r$ is approximately an estimate of $r_{\gamma} \equiv (d - 1)(1 - \gamma)/d$. That is, $r = r_{\gamma}  + \delta_r$ with $\delta_r \ll 1$ a small correction factor. Fig.~\ref{Fig:scaling} demonstrates this for the gateset of Example 1.

To our knowledge, this is the first proof that the RB decay curve is guaranteed to always be exponential for small errors that can be described by CPTP maps -- including gate-dependent errors. This indicates that the model $P_m = A + (B + Cm)p^m$ is not necessary.  Fitting it should always yield $\hat{C} \approx 0$, so estimating $C$ is not likely to help quantify gate-dependence (see suggestion in Refs.~\cite{magesan2012characterizing,magesan2011scalable}).  Instead, our results show that significant non-exponential decay is a clear symptom of non-Markovianity (e.g., time dependence).

We now return to a question raised earlier: Are there natural representations of the perfect and imperfect gatesets in which $\epsilon = r$? ``Natural'' is important, because $\epsilon$ varies so widely over representations. An absurd answer would be to compute $r$ and then search over \emph{all} representations of a gateset to find one in which $\epsilon = r$. The most obvious reasonable option is to arbitrarily fix a CPTP representation of the perfect gateset and to choose the representation of the imperfect gateset in which the gates are all CPTP and $\epsilon$ is minimal ($\epsilon$ can always be made large by choosing a ``bad'' representation -- see Example 2).  This defines a new and gauge-invariant AGsI $\epsilon_{\min} := \min_{M} [\epsilon(\tilde{\mathcal{C}}(M), \mathcal{C})]$, with the minimization restricted such that the gates in $\tilde{\mathcal{C}}(M)$ are CPTP.  But $\epsilon_{\min}$ does \emph{not} exactly correspond to $r$, as it can be strictly \emph{less} than $r$ (see the appendix).

After the initial version of this Letter appeared, Wallman \cite{wallman2017randomized} published an independent analysis of RB. Based on a different representation of the $\mathscr{L}$ operator, Wallman's theory also derives an
exponential decay at the same rate $\gamma$ derived here, but proves a tighter error bound that decays exponentially
with $m$, confirming that the RB decay is completely described by $\gamma$, and $\delta_r$ is negligible in
$r = r_{\gamma} + \delta_r$. Wallman's construction implies that there exists a representation of the imperfect gates for which $\epsilon = r$. To prove this, let $\mathscr{L}'(\mathcal{E}) = \text{avg}_i[\tilde{C}_{i}  \mathcal{E} C_{i}^{-1}]$. $\mathscr{L}'$ has the same spectrum as $\mathscr{L}$. Wallman \cite{wallman2017randomized} gives an explicit construction of a superoperator $\mathcal{L}$ that satisfies $\mathscr{L}'(\mathcal{L}) = \mathcal{L}\mathcal{D}_{\gamma}$, where $\mathcal{D}_{\lambda}$ is a depolarizing channel ($\mathcal{D}_{\lambda} (\rho) = (1 - \lambda)1/d + \lambda\rho$). Now, consider the particular representation of the imperfect Cliffords $\tilde{\mathcal{C}}(\mathcal{L}^{-1}) = \{\mathcal{L}^{-1} \tilde{C}_{i} \mathcal{L}\}$. Some simple algebra (see the appendix) shows that $ r_{\gamma} = \epsilon(\tilde{\mathcal{C}}(\mathcal{L}^{-1}), \mathcal{C})$. So there is an explicitly calculable representation of the gateset that makes $\epsilon = r$. However, the gates in this representation are not generally completely positive, which makes it hard to consider this gauge ``natural" (non-CP gauge choices can even make $\epsilon < 0$).

\vspace{0.1cm}
\noindent\textbf{Conclusions:} It is surprisingly nontrivial to relate the RB error rate $r$ -- a well-defined, representation-independent property of a physical QIP's gates -- to the process matrices describing those gates, and identify what property it corresponds to.  The simple relationship for gate-\emph{independent} errors, where $r$ equals the average gateset infidelity (AGsI, $\epsilon$) between imperfect and perfect Cliffords, obscures the complexity of the general case.  AGsI can be orders of magnitude larger than $r$ unless the right representations are used. This has serious practical consequences, as shown by Example 1 and some of the results in Ref. \cite{epstein2014investigating}, where $r \ll \epsilon$ for experimentally plausible error models.

Our analysis indicates that RB is even more stable and reliable than indicated by previous work \cite{epstein2014investigating,magesan2011scalable,magesan2012characterizing}; $P_m$ decays exponentially (without higher-order corrections of the form $mp^m$) for all small errors describable by process matrices, including coherent errors.  We established this by introducing a new, accurate, theory for the RB decay curve that associates $r$ with a calculable, representation-independent property of the physical gateset.  Subsequent results by Wallman \cite{wallman2017randomized} allow us to observe that this quantity \emph{is} an AGsI, for at least one representation of the imperfect gates, but in this representation the gate process matrices are generally unphysical (not completely positive). Since current theories for many extended RB protocols, such as interleaved \cite{magesan2012efficient}, dihedral \cite{carignan2015characterizing,cross2016scalable}, and unitarity \cite{wallman2014randomized} RB, rely on representation-dependent techniques, it is an interesting open question whether they can be reformulated in a representation-independent way as we did here with basic RB.

\vspace{0.1cm}
\noindent\textbf{Acknowledgements:} 
We acknowledge helpful discussions with Joseph Emerson, Joel Wallman, Steven Flammia, and Marcus da Silva. Sandia National Laboratories is a multimission laboratory managed and operated by National Technology and Engineering Solutions of Sandia, LLC, a wholly owned subsidiary of Honeywell International, Inc., for the U.S. Department of Energy's National Nuclear Security Administration under contract DE-NA0003525. This research was funded, in part, by the Office of the Director of National Intelligence (ODNI), Intelligence Advanced Research Projects Activity (IARPA). All statements of fact, opinion or conclusions contained herein are those of the authors and should not be construed as representing the official views or policies of IARPA, the ODNI, or the U.S. Government.
\bibliographystyle{apsrev}
\bibliography{../../../../../Paper-Library/Bibliography} 

\appendix

\section{The RB protocol}
Here we provide a detailed review of the basic RB procedure, as used throughout the main text. This was defined in Refs.~\cite{magesan2012characterizing,magesan2011scalable}, and see also Refs.~\cite{wallman2014randomized,epstein2014investigating} for detailed descriptions of basic RB. As in the main text, we denote representations of the physical and ideal Cliffords by $\tilde{\mathcal{C}}=\{\tilde{C}_i\}$ and $\mathcal{C}=\{C_i\}$,  respectively, for $i \in [1,2,\dots,|\mathcal{C}|]$. Furthermore, we assume that we can prepare our system in a state $\rho \approx \ket{\psi}\bra{\psi}$ for some pure state $\psi$, and that we can approximately measure whether the system is still in this state, which is represented by the measurement $\mathcal{M} = \{E, \Id-E\}$ where $E \approx \ket{\psi}\bra{\psi}$.

Given some finite set of positive integers $\mathbb{M}$, some $K : \mathbb{M} \to \mathbb{N}$, and some $R \in \mathbb{N}$ (how these quantities are chosen is discussed below), the basic RB protocol is the following:
\begin{enumerate}
\item For each $m \in \mathbb{M}$, pick $K(m)$ sequences of length $m+1$ uniformly sampled from the sub-set of all sequences $\bs{s} \in [1,\dots,|\mathcal{C}|]^{m+1}$ with $C_{s_{m+1}}C_{s_m}\dots C_{s_1} = \Id$.

\item For each sampled sequence $\bs{s}$, apply $\mathcal{S}_{\bs{s},m} =\tilde{C}_{s_{m+1}}\tilde{C}_{s_m}\dots \tilde{C}_{s_1}$ to the initial state $\rho$, and measure to see whether the state has ``survived'' (the measurement $\mathcal{M}$).  Repeat this experiment $R$ times to estimate the \emph{survival probability} $P_{m,\bs{s}} =\text{Tr}(E\mathcal{S}_{\bs{s},m} (\rho))$. We denote this estimate by $\hat{P}_{m,\bs{s}}$.

\item For each $m \in \mathbb{M}$, calculate $\hat{P}_m = \text{avg}_{\bs{s}}[ \hat{P}_{m,\bs{s}}]$, where this average is over the $K(m)$ sequences of length $m+1$ that were sampled. This is an estimate of the average survival probability over \emph{all} possible RB sequences of length $m+1$, which we denote $P_m$.
\end{enumerate}

The experimental estimates for $P_m$ are analyzed by fitting them to the model  \cite{magesan2012characterizing,magesan2011scalable}
\begin{equation}
P_m = A+ (B + Cm)p^m ,
 \end{equation}
where $A$, $B$, $C$, and $p$ are fit parameters. The estimated RB number is obtained from the fit parameters via \cite{magesan2012characterizing,magesan2011scalable}
\begin{equation}
\hat{r} = \frac{d-1}{d} (1-\hat{p}),
\end{equation}
where $d$ is the dimension of the total Hilbert space on which the benchmarking is performed. For the Clifford gates on $n$ qubits, $d=2^n$.

The standard fitting model fixes $C=0$ (this is also called the ``0\textsuperscript{th} order'' fitting model \cite{magesan2012characterizing}). However, Magesan \emph{et al.}~\cite{magesan2012characterizing,magesan2011scalable} suggest that allowing $C\neq 0$ is more appropriate when the error maps might be ``weakly gate-dependent'' (the meaning of this is discussed in the main text), rather than perfectly gate-independent. This is called the ``1\textsuperscript{st} order'' fitting model. Note that the ``1\textsuperscript{st} order'' fitting function given here is different to that presented in Refs.~\cite{magesan2012characterizing,magesan2011scalable}. However, the function herein and the ``1\textsuperscript{st} order'' function given therein have the same functional form (i.e., they are both the sum of a constant term, a purely exponential $p^m$ term, and a $mp^m$ term), which is all that is relevant from the perspective of fitting.

It is clearly essential to make reasonable choices for $\mathbb{M}$ (the set of sequence lengths), $K$ (the function which defines the number of sequences sampled at each length) and $R$ (the number of repeats of each circuit) in order to obtain a good estimate of $r$. How to do this is a statistics problem which we are not concerned with in this Letter. See Refs.~\cite{epstein2014investigating,granade2015accelerated,wallman2014randomized,helsen2017multiqubit} for some work on this.

\subsection*{RB Simulations}
 In this work, we are interested in the underlying property of the gateset which RB estimates. We are \emph{not} interested in how well this property is estimated for physically reasonable choices for $R$, $K$, and $\mathbb{M}$ (see above for the meaning of these quantities). Hence, for all the RB simulations in this paper there is no sampling error on individual experiments (i.e., equivalent to $R \to \infty$), we used $K(m)=500$ for all $m$, and sequence lengths including at least $m \in \{1, 1+50,1+100,\dots,1+2000\}$ (in some simulation we used smaller step sizes between $m$ values). This is to minimize statistical contributions to any deviations of the RB number from the behavior predicted by RB theory. Larger values for the maximum sequence length and greater fixed $K(m)$ (or increasing $K(m)$ with $m$) do not appear to make a significant difference to the results, as is expected \cite{epstein2014investigating}. 

To fit the simulated RB data we use unweighted least squares minimization, and the fit is to the ``1\textsuperscript{st}  order'' model (see above, or main text). This is because Refs.~\cite{magesan2012characterizing,magesan2011scalable} suggest that the ``1\textsuperscript{st} order'' model is more appropriate than the standard fitting model when there could be any gate-dependence to the error maps. However,  for all simulations herein (where the gatesets do have gate-dependent error maps) we found empirically that the RB numbers obtained from the standard fitting model are always within error bars (see below) of the RB numbers obtained from fitting to the ``1\textsuperscript{st} order'' model. This is not surprising, because in the main text we show that the RB decay has a functional form that is well described by $P_m = A+Bp^m$ for any low-error gateset. 

To calculate error bars for estimated RB numbers, the RB estimation protocol was simulated 50 times. The reported estimated RB number is taken to be the average of these estimated RB numbers, and the error on the estimated RB number is taken to be the standard deviation of these estimated RB numbers.

In all simulations, state preparation and measurement (SPAM) are taken to be ideal, with preparation and projection onto the $+1$ eigenstate of $\sigma_z$. Although this is not physically realistic, we have chosen this for conceptual simplicity, as our results are independent of whether or not the SPAM is perfect. Imperfect SPAM affects the asymptotic RB survival probability and the $m=0$ intercept, but not $r$ \cite{magesan2012characterizing,magesan2011scalable}.

In the main text we presented RB decays for Cliffords compiled from imperfect implementations of the $G_x$ and $G_y$ gates. The general behavior of the RB decay for the particular gateset we considered (see main text, Example 1) does not depend upon the details of the compilation table. The particular Clifford compilation table we used is that given in Epstein \emph{et al.}~\cite{epstein2014investigating} (see Table I therein), where we have further decomposed the three non-trivial rotations around the $\sigma_x$ and $\sigma_y$ axes in the obvious way. That is, an $n \pi/2$ rotation around either axis is compiled via $n$ sequential $\pi/2$ rotations, for $n=1,2,3$. We have implemented the identity Clifford gate as ``skip to the next gate'' (the alternatives being to compile the identity Clifford into $G_x$ and $G_y$ gates, or to also include a, possibly imperfect, ``idle'' gate).

In the main text we compare simulated RB decay curves to the decay curves predicted by the ``1\textsuperscript{st} order'' theory given in \cite{magesan2012characterizing,magesan2011scalable}, which is the most accurate RB theory given therein. The ``1\textsuperscript{st} order'' theory decay curve is simply a function of the gateset $\{\tilde{C}_i\}$, and the SPAM operations $\rho$ and $E$. We calculate it directly from the formulas given in Ref.~\cite{magesan2012characterizing}.

\begin{figure}
\begin{overpic}[width=8.5cm]{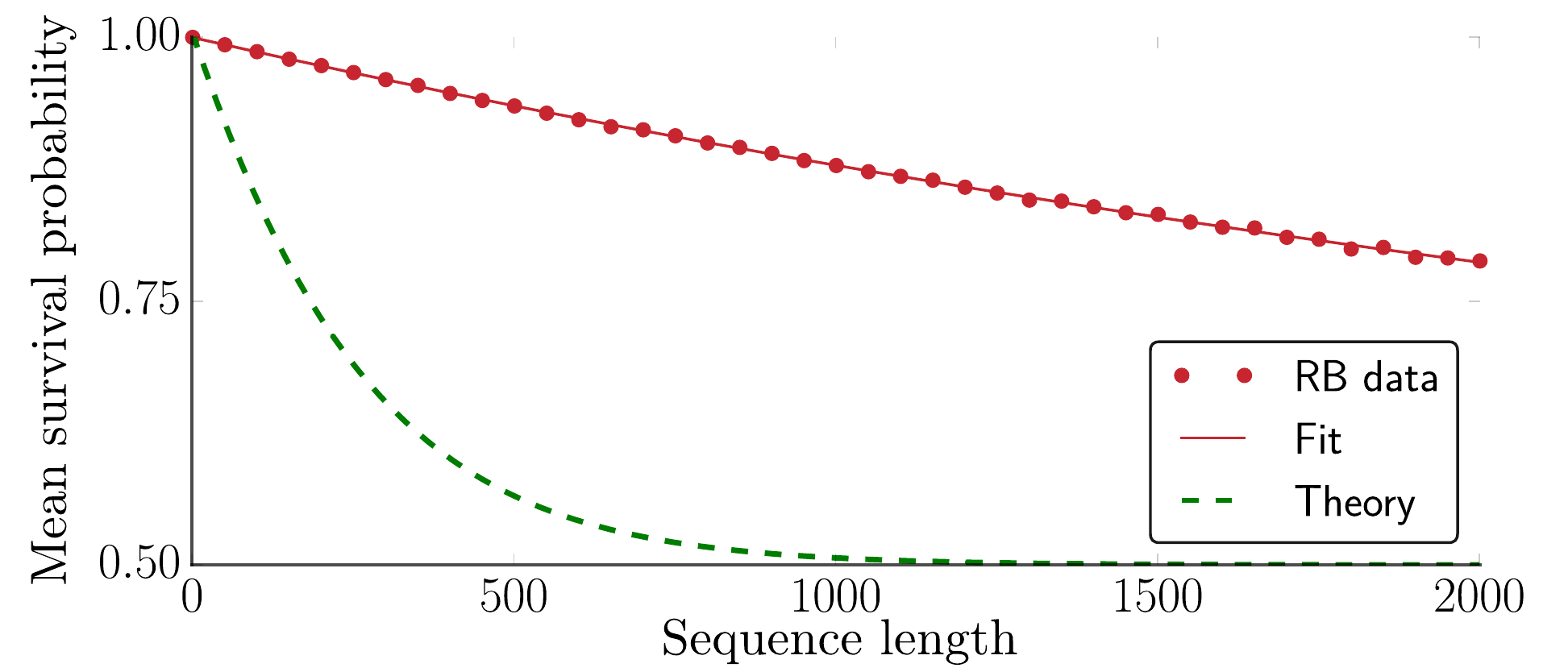}
\end{overpic}
\caption{A simulated RB experiment using Clifford gates compiled from the imperfect implementations of $G_x$ and $G_y$ gates given in Eqs.~(\ref{eq:X-gen} -- \ref{eq:Y-gen}), for the particular case of $\theta_x \hat{v}_x \approx (0.01,0.05,1)\times 10^{-1}$ and $\theta_y \hat{v}_y \approx (0.04,0.03,1) \times 10^{-1}$ and $\lambda = 1 -
( 5 \times 10^{-5})$. There is a clear discrepancy between the prediction of standard RB theory \cite{magesan2012characterizing,magesan2011scalable,wallman2014randomized,epstein2014investigating} and the simulated RB data. This translates into a significant difference between $\hat{r}$ and $\epsilon$, as $\hat{r}=(1.36 \pm 0.03) \times 10^{-4}$ and $\epsilon \approx 2.7 \times 10^{-3}$.}
\label{fig:example-z-error-decay-2}
\end{figure}

\subsection*{Additional examples of $r \ll \epsilon$}
In the main text (see Example 1) we considered the behavior of RB for 1-qubit Cliffords compiled into imperfect gates that are represented by $G_x =  R(\sigma_x,\pi/2)$ and $G_y = R(\sigma_y,\pi/2)$, where 
\begin{equation}
R(H,\theta)[\rho] := \exp(- i \theta H/2) \rho\exp(i \theta H/2).
\end{equation}
 We considered the particular imperfect primitives $\tilde{G}_{x} = R(\sigma_z,\theta)G_x$ and $\tilde{G}_y = R(\sigma_z,\theta)G_y$ with $\theta \ll 1$. Here we show that the discrepancy between the $r$ and $\epsilon$ extends to more general unitary and stochastic errors.  Consider the two imperfect primitives
\begin{align}
\tilde{G}_{x} &= \mathcal{D}_{\lambda}  R(\hat{v}_x \cdot \vec{\sigma},\theta_x)G_x, \label{eq:X-gen}
\\ \tilde{G}_{y} &=  \mathcal{D}_{\lambda}  R(\hat{v}_y \cdot \vec{\sigma},\theta_y) G_y, \label{eq:Y-gen}
\end{align}
for some $\theta_x$, $\theta_y$, $\lambda$, and unit vectors $\hat{v}_x , \hat{v}_y$, which are not necessarily along the $x$ and $y$ axes, where $ \mathcal{D}_{\lambda} = (1 - \lambda) \mathds{1}/d + \lambda \rho$ is a depolarizing channel. These gates have both coherent and stochastic errors. There are a range of value for these parameters for which $\epsilon \gg r$. An example is given in Figure~\ref{fig:example-z-error-decay-2}. However, note that we do not have $\epsilon \gg r$ for all values of these parameters.

\subsection*{The gauge-minimized AGsI can be smaller than $r$}
Consider an imperfect gateset that may be represented by the CPTP maps $\tilde{\mathcal{C}}_a = \{\tilde{C}_{i,a} \}$ where
\begin{equation}
\tilde{C}_{i,a}  = \mathcal{D}_{\lambda} C_i ,
\end{equation}
with $\mathcal{C} = \{C_i\}$ the standard representation of the 1-qubit Clifford gates (i.e., the automorphism group of the Pauli matrices), and $\mathcal{D}_{\lambda}(\rho) = (1 - \lambda) \mathds{1}/d + \lambda \rho$ with $1 \geq \lambda \geq -1/3$, which is a uniform depolarization channel. For this gateset, $r =  1 - \bar{F}(\mathcal{D}_{\lambda},\mathds{1}) = \epsilon(\tilde{\mathcal{C}}_{a},\mathcal{C})$, as the error maps are gate independent (see main text, or Refs.~\cite{magesan2011scalable,magesan2012characterizing}). Now, for $\alpha > 0$, define the invertible linear map $M_{\alpha}$ by
\begin{align}  
M_{\alpha}(\mathds{1}) &= \mathds{1},\hspace{0.01cm} &M_{\alpha}(\sigma_x) &= \sigma_x, \\
M_{\alpha}(\sigma_y) &= \alpha \sigma_y ,\hspace{0.01cm} &M_{\alpha}(\sigma_z) &= \sigma_z.
\end{align}
In terms of this map, define the gateset representation $\tilde{\mathcal{C}}_b = \{\tilde{C}_{i,b} \}$ where
\begin{equation}
 \tilde{C}_{i,b} =M_{\alpha} \tilde{C}_{i,a}M_{\alpha}^{-1}.
\end{equation}
By construction, $\tilde{\mathcal{C}}_b$ is gauge-equivalent to $\tilde{\mathcal{C}}_a$. So both $\tilde{\mathcal{C}}_a$ and $\tilde{\mathcal{C}}_b$ can represent the same physical gateset. As such, these representations are associated with the same $r$, and we know that $r = \epsilon(\tilde{\mathcal{C}}_a,\mathcal{C})$ from above. We now show that for any non-trivial depolarizing channel (i.e., $\lambda <1$) there exists a range of values for $\alpha$ such that 
\begin{enumerate}
\item $ \epsilon(\tilde{\mathcal{C}}_b,\mathcal{C}) < \epsilon(\tilde{\mathcal{C}}_a,\mathcal{C})= r$.
\item All of the gates in $\tilde{\mathcal{C}}_b$ are CPTP. 
\end{enumerate}
This then implies that the AGsI to the target gates of the CPTP representation of the physical gateset that has the minimal AGsI to the targets is smaller than $r$ for this gateset. That is, $\epsilon_{\min} < r$, in this case, where $\epsilon_{\min}$ is defined in the main text.

Using the same notation as for the AGsI of a gateset, denote the AGI of a gate representation $\tilde{G}$ to a target $G$ by $\epsilon(\tilde{G},G) \equiv 1 - \bar{F} (\tilde{G},G)$. Using the relations from Refs.~\cite{dugas2016efficiently,wallman2015estimating} (e.g., see Eq.~(10) in Ref.~\cite{dugas2016efficiently}), the AGI of a trace-preserving map $\tilde{G}$ to a target $G$, may be written as 
\begin{equation}
\epsilon(\tilde{G},G)= \frac{d^2- \text{Tr}(\Lambda)}{d(d+1)},
\label{Eq:AGI-as-trace}
\end{equation}
 where $d$ is the dimension of the Hilbert space, and $\Lambda = \tilde{G}G^{-1}$. The depolarizing channel $\mathcal{D}_{\lambda}$ commutes with $M_{\alpha}$, and so the AGI of $\tilde{C}_{i,b}$ to the target Clifford $C_{i}$ is 
\begin{equation}
\epsilon(\tilde{C}_{i,b},C_{i})= \frac{1}{6}(4- \text{Tr}(\Lambda_{i,b})),
\label{Eq:AGI-as-trace2}
\end{equation}
 where 
\begin{equation}
\Lambda_{i,b} \equiv  \tilde{C}_{i,b} C^{-1}_{i} = \mathcal{D}_{\lambda} M_{\alpha}C_{i} M_{\alpha}^{-1}C^{-1}_i.
\end{equation}

For any $i$ for which the Clifford $C_i$ maps $\sigma_y \to \sigma_y$ up to phase, $\Lambda_{i,b} = \mathcal{D}_{\lambda}$, and hence $\tilde{C}_{i,a}$ and $\tilde{C}_{i,b}$ have the same AGI to $C_i$. From the definition of $M_{\alpha}$ and because Cliffords map Pauli operators to Pauli operators, for any $i$ for which the Clifford gate $C_i$ does \emph{not} map $\sigma_y \to \sigma_y$, up to phase, it follows that
\begin{align}
 \Lambda_{i,b}(\mathds{1}) &= \mathds{1}, \hspace{0.1cm} &\Lambda_{i,b}(\sigma_y) &= \lambda \alpha \sigma_y, \label{Eq:error-maps-i1}\\
\Lambda_{i,b}(\sigma_l) &= \frac{\lambda}{\alpha}\sigma_l,\hspace{0.1cm} & \Lambda_{i,b}(\sigma_m)& = \lambda \sigma_m,
\label{Eq:error-maps-i}
\end{align}
where $l$ and $m$ are some ordering of $x$ and $z$ (the exact labelling depends on $i$, but is irrelevant here). Therefore, from Eq.~(\ref{Eq:AGI-as-trace2}) the AGI of $\tilde{C}_{i,b}$ to $C_{i}$ for any such $i$ is
\begin{equation}
\epsilon(\tilde{C}_{i,b},C_{i})= \frac{1}{6}\left(3 - \lambda \frac{\alpha^2 + \alpha + 1}{\alpha}\right).
\end{equation}

For any $\alpha >0$ with $\alpha \neq 1$, this AGI is smaller than $\epsilon(\tilde{C}_{i,a},C_{i})$ (which is given by taking $\alpha=1$ in this equation). Therefore, for all $\alpha > 0$ except $\alpha = 1$, the AGsI, which is simply the average of the AGIs of the gates in the gateset, of $\tilde{\mathcal{C}}_{b}$ to $\mathcal{C}$ is smaller than the AGsI of $\tilde{\mathcal{C}}_{a}$ to $\mathcal{C}$. That is, $\epsilon(\tilde{\mathcal{C}}_b,\mathcal{C}) < \epsilon(\tilde{\mathcal{C}}_a,\mathcal{C})$. Note that for general $\alpha$ it is possible that $\epsilon(\tilde{\mathcal{C}}_b,\mathcal{C}) <0$, which is because the gates in $\tilde{\mathcal{C}}_{b}$ are not all completely positive for all values of $\alpha$.

We now confirm that for any $\lambda <1$ there are values of $\alpha \neq 1$ such that $\tilde{\mathcal{C}}_{b}$ is CPTP. All of the gates in $\tilde{\mathcal{C}}_{b}$ are obviously TP. The map $\tilde{C}_{i,b}$ is CP if the error map $\Lambda_{i,b}$ is CP. A map is CP if all of the eigenvalues of the Choi matrix are non-negative \cite{skowronek2012choi,choi1975completely}, where the Choi matrix $\chi$ for a map $G$ is defined by
\begin{equation}
\chi(G) = \sum_{i,j=1}^d B_{ij} \otimes G(B_{ij}),
\end{equation}
with $B_{ij}$ the $d\times d$ matrix with 1 in the $ij$-th entry and 0s elsewhere (the standard basis for matrices). For those gates for which the error map is simply $ \mathcal{D}_{\lambda}$, then the error map is clearly CP as this is a depolarizing channel. Hence, we need only consider those gates for which the error maps are given by Eq.~(\ref{Eq:error-maps-i1} -- \ref{Eq:error-maps-i}). 
For any such error map, the eigenvalues of its Choi matrix, denoted $\xi_{i}$ with $i=0,1,2,3$, are given by
\begin{equation}
\xi_{j+2k} = (-1)^j\lambda\alpha^2 +(1+(-1)^{k+j}\lambda)\alpha + (-1)^{k} \lambda.
\end{equation}
For any non-identity depolarizing channel (so $\lambda < 1$) there exists an $\alpha \neq 1$ such that all of these eigenvalues are positive (and hence all the gates are CP), which may be easily confirmed numerically. As such, there exists values of $\alpha$ for any non-trivial $\lambda$ such that (1) $ \epsilon(\tilde{\mathcal{C}}_b,\mathcal{C}) < \epsilon(\tilde{\mathcal{C}}_a,\mathcal{C}) = r$, and (2) all of the gates in $\tilde{\mathcal{C}}_b$ are CPTP. In turn, this implies that $\epsilon_{\min} < r$ for at least some gatesets.

\subsection*{Error bounds for the approximate RB theory}
In the main text, we started from the exact average survival probability over all RB sequences of length $m$, which is given by
\begin{equation}
P_m = \frac{1}{|\mathcal{C}|^m}\sum_{\bs{s}}  \text{Tr} (E \tilde{C}_{\bs{s}^{-1}} \tilde{C}_{s_m} \dots \tilde{C}_{s_{1}}  (\rho )) ,
\end{equation}
and by noting that $\tilde{C}_{\bs{s}^{-1}} = (\bar{\Lambda} + \Delta_{\bs{s}^{-1}})  C_{s_1}^{-1} \dots C_{s_m}^{-1} $, where $\Delta_{i} = \Lambda_i - \bar{\Lambda}$, this was rewritten as
\begin{equation*}
P_m = \frac{1}{|\mathcal{C}|^m}\sum_{\bs{s}} \text{Tr} (E \bar{\Lambda}   C_{s_1}^{-1} \dots C_{s_m}^{-1} \tilde{C}_{s_m} \dots \tilde{C}_{s_{1}} (\rho)) +\tilde{\delta}_m.
\end{equation*}
Here $\tilde{\delta}_m$ is the correction required so that this equality holds, and is given explicitly by
\begin{equation}
\tilde{\delta}_m = \frac{1}{|\mathcal{C}|^m}\sum_{\bs{s}} \text{Tr} (E \Delta_{\bs{s}^{-1}}  C_{s_1}^{-1} \dots C_{s_m}^{-1} \tilde{C}_{s_m} \dots \tilde{C}_{s_{1}} (\rho)).
\label{tildedeltam}
\end{equation}
We now prove that
\begin{equation}
|\tilde{\delta}_m| \leq \delta_{\diamond} \equiv \frac{1}{2} \text{avg}_i \|  \Lambda_i -  \bar{\Lambda}\|_{\diamond},
\label{eq-to-prove}
\end{equation}
for all $m$, as claimed in the main text, where $ \|  \cdot \|_{\diamond}$ is the diamond norm.

We begin by defining the diamond norm. Let $L(\mathcal{H})$ denote the space of linear operators on some Hilbert space $\mathcal{H}$. For $\rho \in L(\mathcal{H})$, define $\|\rho\|_1 := \text{Tr}(\sqrt{\rho^{\dagger}\rho})$. For a superoperator $\mathcal{A} : L(\mathcal{H}) \to  L(\mathcal{H}) $, the diamond norm is defined by \cite{watrous2005notes,lidar2013quantum,aharonov1998quantum}
\begin{equation}
\| \mathcal{A} \|_{\diamond} := \sup_{\rho} \| [\mathcal{A} \otimes \mathds{1}](\rho) \|_1 ,
\end{equation}
where $\mathds{1}$ is the identity superoperator on $L(\mathcal{H})$, and the supremum is over all $\rho \in L(\mathcal{H} \otimes \mathcal{H})$ with $\| \rho \|_1 = 1$.  The diamond norm has the following properties: For any CPTP maps $\mathcal{A}$ and $\mathcal{B}$ on $L(\mathcal{H})$, for any linear maps $\mathcal{A}'$ and $\mathcal{B}'$ on $L(\mathcal{H})$, and any density operator $\rho$ and measurement effect $E$ on $\mathcal{H}$, we have
\begin{align}
\| \mathcal{A} \|_{\diamond} &= 1,\label{d1} \\
\| \mathcal{A}'\mathcal{B}' \|_{\diamond} &\leq \|\mathcal{A}'\|_{\diamond}\|\mathcal{B}'\|_{\diamond},\label{d2} \\
2|\text{Tr}(E [\mathcal{A} -\mathcal{B}](\rho)) |&\leq \| \mathcal{A}-\mathcal{B} \|_{\diamond}.\label{d3}
\end{align}

The first of these properties follows easily from the definition of the diamond norm (see also Ref.~\cite{aharonov1998quantum}). The second property is proven in Ref.~\cite{aharonov1998quantum}. The final property can be proven in the following way: for any $\mathcal{A}$, $\mathcal{B}$, $\rho$, and $E$ as above, then
\begin{align}
\| \mathcal{A} -\mathcal{B} \|_{\diamond} &\geq \|  [\mathcal{A} -\mathcal{B}](\rho)\|_1, \vphantom{\max_T} \\
&= 2 \max_{T \leq \mathds{1}} \left[ \text{Tr}(T [\mathcal{A} -\mathcal{B}](\rho) ) \right],\label{max-Tr} \\
&\geq 2 |  \text{Tr}(E [\mathcal{A} -\mathcal{B}](\rho) )| \vphantom{\max_T},
\end{align}
where the maximization in Eq.~(\ref{max-Tr}) is over all positive operators $T$ satisfying $T\leq \mathds{1}$, and this equality follows from the relation   $ \|  \sigma - \sigma' \|_1 = 2 \max_{T \leq \mathds{1}} \left[ \text{Tr}(T (\sigma - \sigma')) \right]$, for any density operators $\sigma$ and $\sigma'$, which is given in Refs.~\cite{gilchrist2005distance,nielsen2010quantum}.

We are now ready to prove Eq.~\eqref{eq-to-prove}. Using the properties of the diamond norm given above, we have that:
\begin{widetext}
\begin{align} 
|\tilde{\delta}_m| & \leq  \frac{1}{|\mathcal{C}|^m}\sum_{\bs{s}} |\text{Tr} (E (\Lambda_{\bs{s}^{-1}} - \bar{\Lambda})  C_{s_1}^{-1} \dots C_{s_m}^{-1} \tilde{C}_{s_m} \dots \tilde{C}_{s_{1}} (\rho))|,\label{ne1} \\
&\leq   \frac{1}{2|\mathcal{C}|^m}\sum_{\bs{s}} \| (\Lambda_{\bs{s}^{-1}} - \bar{\Lambda})  C_{s_1}^{-1} \dots C_{s_m}^{-1} \tilde{C}_{s_m} \dots \tilde{C}_{s_{1}}\|_{\diamond}, \label{ne2}\\
&\leq   \frac{1}{2|\mathcal{C}|^m}\sum_{\bs{s}} \| (\Lambda_{\bs{s}^{-1}}- \bar{\Lambda}) \|_{\diamond}\| C_{s_1}^{-1} \|_{\diamond} \dots \| C_{s_m}^{-1}  \|_{\diamond}\|\tilde{C}_{s_m} \|_{\diamond} \dots  \| \tilde{C}_{s_{1}}\|_{\diamond}, \label{ne3}\\
&\leq   \frac{1}{2|\mathcal{C}|^m}\sum_{\bs{s}} \| (\Lambda_{\bs{s}^{-1}} - \bar{\Lambda}) \|_{\diamond}, \label{ne4}\\
&=  \frac{1}{2|\mathcal{C}|}\sum_{i=1}^{|\mathcal{C}|} \| (\Lambda_i - \bar{\Lambda}) \|_{\diamond}, \label{ne5}\\
& = \delta_{\diamond} \label{ne6}.
\end{align} 
\end{widetext}
Eq.~(\ref{ne1}) follows from Eqs.~(\ref{tildedeltam}) and the triangle inequality; Eq.~(\ref{ne2}) follows from Eq.~(\ref{d3}); Eq.~(\ref{ne3}) follows from Eq.~(\ref{d2}); Eq.~(\ref{ne4}) follows from Eq.~(\ref{d1}); Eq.~(\ref{ne5}) follows by noting that there are $|\mathcal{C}|^{m}$ terms which are being summed over, in the quantity of Eq.~(\ref{ne4}), and $1/|\mathcal{C}|$ of them are associated with each possible inversion Clifford; Eq.~(\ref{ne6}) simply follows from the definition for $\delta_{\diamond}$. This concludes our proof that  $|\tilde{\delta}_m| \leq  \delta_{\diamond} $.

\subsection*{Relating $r$ to an AGsI}
Following the main text and Wallman \cite{wallman2017randomized}, let $\mathscr{L}'(\mathcal{E}) = \text{avg}_i[ \tilde{C}_{i}  \mathcal{E} C_{i}^{-1}]$ \cite{wallman2017randomized}, and observe that $\mathscr{L}'$ has the same spectrum as $\mathscr{L}$. Wallman \cite{wallman2017randomized} gives an explicit construction, in terms of the imperfect Clifford superoperators $\tilde{\mathcal{C}}$, of a superoperator $\mathcal{L}$ that satisfies $\mathscr{L}'(\mathcal{L}) = \mathcal{L}\mathcal{D}_{\gamma}$. Now, consider the particular gauge representation of the imperfect Cliffords given by $\tilde{\mathcal{C}}(\mathcal{L}^{-1}) = \{\mathcal{L}^{-1} \tilde{C}_{i} \mathcal{L}\}$. As long as $\mathcal{L}$ is invertible, then we have $\bar{\Lambda}_{\mathcal{L}} = D_{\gamma}$ where $\bar{\Lambda}_{\mathcal{L}} \equiv \text{avg}_i[ \mathcal{L}^{-1} \tilde{C}_{i} \mathcal{L} C_{i}^{-1}]$. Taking the AGI to the identity of both sides of this equation obtains 
\begin{equation}
\epsilon(\bar{\Lambda}_{\mathcal{L}},\mathds{1}) = \frac{(d-1)(1-\gamma)}{d} \equiv r_{\gamma}.
\end{equation}

 Hence, as we have already argued in the main text that $r = r_{\gamma}+ \delta_r$ with $\delta_r$ negligible, we have that $r = \epsilon(\bar{\Lambda}_{\mathcal{L}},\mathds{1}) + \delta_r$. Now, $\bar{\Lambda}_{\mathcal{L}}$ is simply the average error map calculated in a particular gauge (conjugation of the $\tilde{C}_i$ by $\mathcal{L}^{-1}$ is a gauge transformation). In particular, it then immediately follows that $r =\epsilon(\tilde{\mathcal{C}}(\mathcal{L}^{-1}), \mathcal{C}) +\delta_r$. Hence, $r \approx \epsilon(\tilde{\mathcal{C}}(\mathcal{L}^{-1}), \mathcal{C})$ with the approximation error negligible. 
\end{document}